# Quantum Optimization Algorithms in Operations Research: Methods, Applications, and Implications

Florian Klug
Munich University of Applied Sciences, florian.klug@hm.edu

**Quantum optimization algorithms (QOAs) have the potential to fundamentally transform the application of optimization methods in decision making. For certain classes of optimization problems, it is widely believed that QOA enables significant run-time performance benefits over current state-of-the-art solutions. With the latest progress on building quantum computers entering the industrialization stage, quantum-based optimization algorithms have become more relevant. The recent extreme increase in the number of publications in the field of QOA demonstrates the growing importance of the topic in both the academia and the industry. The objectives of this paper are as follows: (1) First, we provide insight into the main techniques of quantum-based optimization algorithms for decision making. (2) We describe and compare the two basic classes of adiabatic and gate-based optimization algorithms and argue their potentials and limitations. (3) Herein, we also investigate the key operations research application areas that are expected to be considerably impacted by the use of QOA in decision making in the future. (4) Finally, current implications arising from the future use of QOA from an operations research perspective are discussed.**

**1. INTRODUCTION**

Quantum computing based on the principles of quantum mechanical phenomena has seen substantial advancements over the last few years. The ongoing evolution of quantum computing hardware and the advances of quantum algorithms have already demonstrated for specific problems, quantum speedups over classical computing (Arute et al., 2019). The intrinsic parallelism, based on the superposition principles of quantum states, in combination with interference effects, lead to exponential computing capabilities, which offer interesting possibilities in operations research (Creemers & Perez, 2022). From an optimization perspective, the superposition of quantum bits (qubits) – which exist in multiple states simultaneously - is therefore inherently well suited to solve optimization problems. However, the probabilistic nature of qubit measurement requires clever quantum algorithms to boost the probability of measuring some desired state that corresponds to a problem solution. Therefore, superposition alone is not sufficient to achieve a quantum advantage (Symons et al, 2023).

As we progress toward large-scale quantum computers, operations researchers are increasingly addressing quantum-based optimization problems (Parekh, 2023). Modelling and solving decision problems with the help of quantum computing may demonstrate a number of potential performance gains over established optimization algorithms such as the following:

- Existing classical optimization algorithms can be sped up by exploiting specific quantum effects (Sao et al., 2019).[1]
- Some classes of problems can be encoded and solved more efficiently by quantum algorithms than classical algorithms, such as integer factorization problems, unordered search problems, or discrete logarithms (Nannicini, 2020).
- Theory and classical simulations indicate that, for some problems, new quantum algorithms allow more effective approaches to solving certain classes of difficult combinatorial optimization problems (Lubinski et al., 2023).
- New insights into complex algorithms and complexity theory have been found (Bernstein & Vazirani, 1993; Watrous, 2008).

This article provides an up-to-date overview of the main quantum-based optimization techniques and their application in operations research. The remainder of this paper is organized as follows: In the next section, we provide brief insight into the principles of quantum computing to describe QOAs in OR in the following Section 2. After introducing the two basic architectures of quantum computing, annealing-based quantum computing (Section 2.1) and gate-model quantum computing (Section 2.2) with their differences (Section 2.3), future application areas of QOAs in OR are presented (Section 2.4). Section 3 then addresses the resulting main implications for the application of QOA algorithms in OR, which are summarized in the form of five implication statements. The paper concludes in Section 4 with a discussion of future challenges in the application of QOA.

---

[1] Quantum computing also motivates classical theorists to find better classical algorithms.

## 2. QOAs IN OR

QOAs are quantum algorithms used to solve optimization problems, which run on a realistic model of quantum computation. Independent of the multitude of OR application areas, an optimization algorithm aims to find the best solution to a problem from a set of possible options, given its desired outcome and constraints. The advantage of quantum parallelism is used where a superposition of an exponential number of classical states can be considered simultaneously. Supported by quantum annealing (see Section 2.1) or interference effects (see Section 2.2), this quantum state that encodes the problem is transformed to a quantum state that generates a solution. As with probabilistic classical algorithms, reading out the results leads to a probability distribution. However, the main advantage of QOAs lies in the utilization of constructive interference or quantum mechanical fluctuations, which are not accessible to classical algorithms. Nevertheless, the magnitude of available quantum speedups for OR applications is often hard to assess and can be obscured by intricate technical details about the underlying subroutines and their complexities (Dalzell et al., 2023).

Six main OR application areas (see Section 2.4) could be analyzed, which are likely to be strongly influenced by the use of QOA in decision making in the future (see Fig. 1). The challenge from the OR point of view is the development of suitable QOA to make decision problems more efficient and faster in the future with the help of quantum computers (see Section 3). Simply developing a QOA with a theoretical quantum speedup is not sufficient, OR problems must be simultaneously valuable to the user and also difficult to obtain classically (Dalzell et al., 2023).

While the implementation of classical optimization algorithms is largely independent of the computer technology used, the implementation of quantum algorithms depends on the underlying technology. The large number of QOAs already in existence can be attributed to two fundamental different quantum computing architectures: annealing-based and gate-model quantum computing. As for the classical algorithm design, QOAs make extensive use of quantum subroutines, such as quantum Fourier transforms (QFTs), which serve as a central building block for many quantum algorithms. The impossibility of non-intrusive measurement and unitarity on quantum operations (see Section 2.2) limit the number of quantum subroutines (algorithmic paradigms) compared to the number of known classical subroutines (Coles et al., 2018).

The qubit, as the elementary building block of a quantum computer, can be implemented using a variety of physical two-level systems, such as persistent currents in a superconducting circuit, trapped ions, silicon quantum dots,

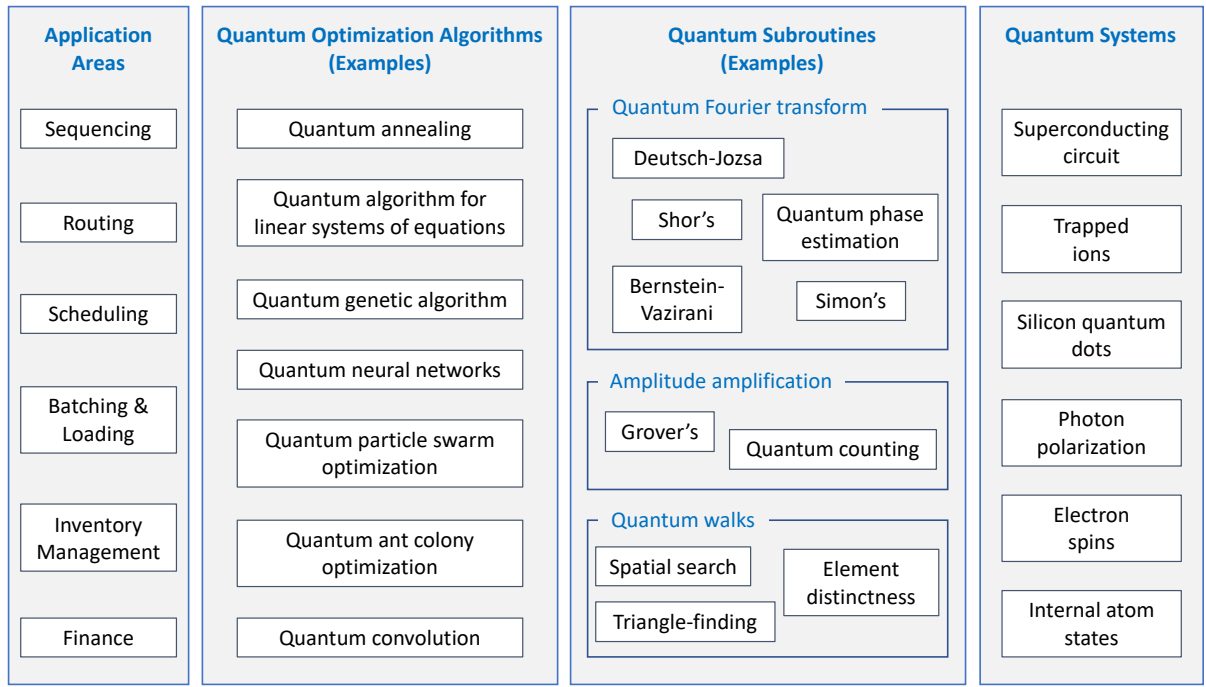

**Fig. 1.** Quantum landscape of optimization algorithms in operations research

photon polarization, electron spins, or internal atom states (see Fig. 1).

**2.1 QUANTUM ANNEALING**

Quantum annealing (QA) is a physically motivated generic metaheuristic that uses quantum–mechanical fluctuations for finding the lowest energy state (ground state) by traversing over the solution space. According to quantum mechanics, the time evolution of the state $\Psi(t)$ is described by the time-dependent Schrödinger equation as follows:

$$H|\Psi(t)\rangle = i\hbar \frac{\partial}{\partial t}|\Psi(t)\rangle \quad (1)$$

where the Hamiltonian H encoding a given problem is an operator describing the energy of the system. Each quantum state $\Psi(t)$ can be written as a linear combination (superposition) of energy eigenstates as follows:

$$\Psi(t) = \sum E_i |e_i\rangle \quad (2)$$

Each quantum system has discrete stationary energy levels that represent solution values, described by a basis of energy eigenstates $\{|e_i\rangle\}_{i=0,1,..}$ and

$$H|e_i\rangle = E_i|e_i\rangle \quad (3)$$

with the minimum energy state $|e_0\rangle$ and energy eigenvalue $E_0$ (ground state) (see Fig. 2).

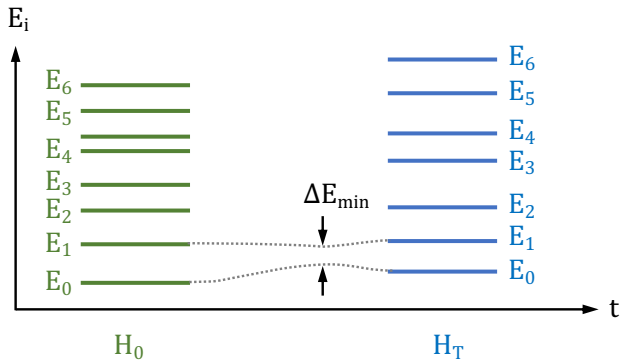

**Fig. 2.** Adiabatic evolution of energy states (Lobe, 2021)

The natural tendency of physical systems to reach the minimum energy states in both the macroscopic and the microscopic world, is used to generate the solution of a combinatorial optimization problem with a discrete search space. According to the adiabatic theorem, a system that is in an eigenstate of the known initial Hamiltonian $H_0$ will very likely end up in the corresponding eigenstate of the unknown target Hamiltonian $H_T$ if the evolution is sufficiently slow (Farhi et al., 2000). This applies not only to the lowest state $E_0$ but also to every energy eigenstate and therefore preserves the energy ranking shown in Fig. 2. The adiabatic evolution of energy states can be used to minimize a function H(t): $\{0,1\}^n \to \mathbb{R}$ at time *T* after the binary encoding of the possible solutions:

$$H(t) = \left(1 - \frac{t}{T}\right)H_0 + \frac{t}{T}H_T \quad (4)$$

---

[2] Throughout the paper, $\mathcal{O}(f(n))$ is interpreted as asymptotically upper-bound by *f(n)*.

In general, QA starts with an equally distributed superposition of all of the possible n digit quantum states $|x\rangle$ to prepare the simple initial system with a known ground state:

$$|e_0\rangle = \frac{1}{\sqrt{2^n}} \sum_{x \in \{0,1\}^n} |x\rangle \quad (5)$$

This known ground state is then slowly transferred to generate the targeted ground state, representing the optimal solution to the problem. Once the problem Hamiltonian (3) exits, energy states ($E_i > E_0$) are introduced. In the annealing process, quantum fluctuations (tunneling) cause transitions between states (Kadowaki & Nishimori, 1998). The quantum mechanical effect of quantum tunneling enables QOA to search the solution space more efficiently and faster than classical annealing algorithms (Morita & Nishimori, 2008; Rønnow et al., 2014). During the annealing process, the first exiting state $E_1$ approaches the ground state $E_0$ closely (see Fig. 2). This difference $\Delta E_{min}$ between the best and the second-best solution, called the spectral minimum gap, determines the minimal run time (and therefore, the time complexity) for an adiabatic process with $\mathcal{O}(\Delta E_{min}^{-2})^2$. This gap typically becomes smaller when the optimization problem size increases, which enlarges the annealing time. Therefore, the minimum gap is the main parameter to measure the efficiency of the optimization algorithm of a given problem.

QA problems must be converted into an approach that can be understood and processed by the computer hardware. For this purpose, a special case of the Hamiltonian (4) is modeled after the objective function of a quadratic unconstrained binary optimization (QUBO) problem. The unified modeling framework of QUBO can be easily converted to the affine problem of the so-called Ising model, which is more common in a physical context and will therefore not be described further. QUBO plays a central role in solving combinatorial problems in OR, including network flows, facility and resource allocation problems, clustering and set partitioning problems, scheduling and sequencing, max-cut, max-clique, vertex cover, and other graph and management science problems (Glover et al., 2018; Kochenberger et al., 2014).

With the control parameter *k*, defined as $t/T_A$, where $T_A$ represents the total annealing time and *t* denotes the time, the classical QUBO problem is transformed into a problem-specific quantum version of the Hamiltonian with *N* qubits:

$$H(k) = A(k) \sum_{i=1}^{N} a_i q_i + B(k) \sum_{\substack{i,j=1 \\ i<j}}^{N} b_{ij} q_i q_j \qquad (6)$$

QUBO problems can be represented by an interaction graph, where the nodes are $q_i \in \{0,1\}$ with node weight $a_i$ (bias) and edge weight $b_{ij}$ (coupling). The goal is to generate a binary sequence $q_i^*$ that encodes the optimal solution, according to the specific problem set $a_i$ and $b_{ij}$, which represent the constraints of the optimization problem. Two quantum bits $q_i$ and $q_j$ are pairwise coupled with the adjustable strength $b_{ij}$, which controls the influence exerted between two qubits.

The weight $a_i$ (bias) is an adjustable term associated with the qubit's tendency to collapse into its two possible final states. The first term in (6) describes the initial or superposition Hamiltonian, and the second term represents the final or problem Hamiltonian. During the annealing process, the weight $A(k)$ of the initial Hamiltonian is reduced from one to zero, while the weight $B(k)$ of the final Hamiltonian is enlarged from zero to one, staying in the minimum energy state throughout the whole annealing cycle (see Fig. 3).

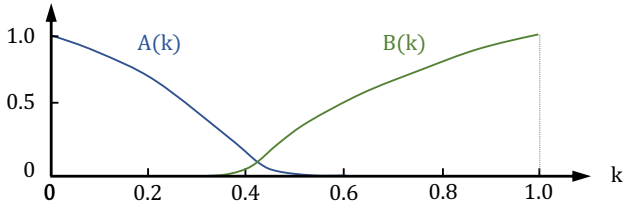

**Fig. 3.** Weight development of the initial (A) and final (B) Hamiltonian

The nodes of the graphical representation are mapped to qubits, and the edges are mapped to the so-called couplers. Qubits are physically realized by superconducting loops with their corresponding magnetic fields, where the circulating currents represent binary quantum states as spin states (0 = current running clockwise, 1 = current running counterclockwise) (Johnson et al., 2011). QA is performed by applying and varying transverse external magnetic fields $\vec{B}$ to the superconducting loops, which are connected to an array with programmable spin-spin couplings. The adiabatic cycle starts with a superposition of the binary quantum states by a large transverse magnetic field (see Fig. 4a). When the problem space is explored and the transition probability is lowered, the bias weight $a_i$ (representing the tendency for the qubit to end up in state 0 or 1) gradually raises the energy barrier between the two possible states of each qubit (see Fig. 4b). In contrast to classical simulated annealing, where the transition probability depends on the height of the energy barrier, quantum fluctuations can penetrate through potential barriers on the basis of the quantum mechanical phenomenon of tunneling. The couplers $b_{ij}$ represent the influence that two qubits $i$ and $j$ have on each other according to the specific optimization problem mapped in (6). The setting of the problem-specific biases $a_i$ and coupling strengths $b_{ij}$ provides a spin-spin coupling energy that allows the spins to favor the alignment or anti-alignment whilst finding the lowest energy configuration (either equal 00, 11 or opposite 01, 10). With the exploration of the problem space, energetically favorable states (set by the Hamiltonian) are chosen, which evolve to the solution of the problem set. During the annealing cycle, the external magnetic field and, therefore, the quantum fluctuations are reduced until a complete quantum freezeout is reached (Morita & Nishimori, 2006). The gradual reduction of the magnetic field plays the same role as that of thermal fluctuations in the conventional simulated annealing approach. At $\vec{B} = 0$, the qubit lattice is relaxed and the quantum annealer outputs a binary sequence $q_i^*$ configuration corresponding to the eigenstate and the eigenenergy (8), which satisfies the optimal solution of the problem with the highest probability (see Fig. 4c).

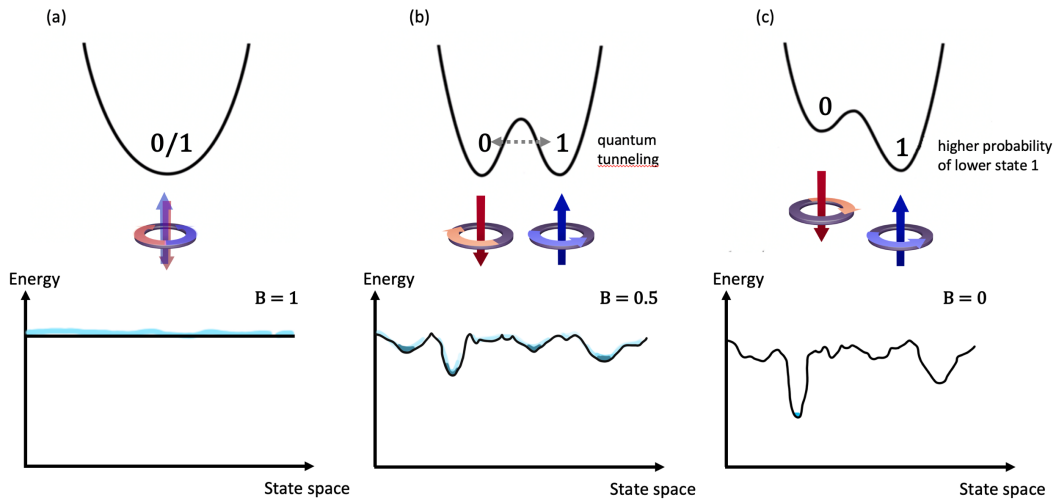

**Fig. 4.** Quantum annealing process

## 2.2 Universal gate-based optimization algorithms

A universal and computationally equivalent alternative to solving optimization problems with quantum annealing is gate-based modeling (Aharonov et al., 2008). As in classical computation, quantum gates can be combined like Boolean gates to achieve quantum computation. Elementary logic gates are used as building blocks to create quantum circuits (Barenco et al., 1995). Gate-based algorithms actively control and manipulate the evolution of quantum states over time, as compared to analog quantum annealing, where the natural tendency of a physical system is used to find the state of minimum energy. The initial state, stored in quantum registers, is processed by a series of quantum gates acting on these quantum registers (see Fig. 5). Besides the quantum version of classical Boolean logic gates (Pauli X-gate is a quantum equivalent of a NOT-gate), there are quantum-specific gates (e.g., Hadamard or controlled-Not gate). It has been proven that any quantum circuit can be approximated using only three gates (Shi, 2002)[3]. Furthermore, the Solovay–Kitaev theorem proves that an arbitrary quantum circuit of $m$ constant-qubits gates can be efficiently approximated using a certain number of gates to an accuracy $\epsilon$ from the order $\mathcal{O}(m \, log^c \, (m/\epsilon))$, where $c$ is a constant approximately equal to 2 (Nielsen & Chuang, 2010).

Quantum logic gates can image all classically known circuits but are reversible; thus, all of the input values can be uniquely determined by their output values. Each quantum gate is mathematically represented by a unitary transformation, describing the evolution of a quantum system with the time evolution operator or propagator $U$ ($U^\dagger = U^{-1}$). Unitary evolutions of the quantum state ensure that the sum of the individual probabilities of all of the possible measurement results (eigenstates) of a quantum system $|\Psi\rangle$ satisfies the normalization condition. Every gate has an equal number of input and output qubits. Reversible gates are also required to maintain the quantum mechanical properties of entanglement and superposition.

In order to retain results, measurements of the quantum system are necessary. As measurement disturbs the superposition state of the quantum system (collapse of the wave function), quantum algorithms use quantum interference to bias the measurement of a qubit toward a desired solution of the problem set. Each quantum algorithm uses a selective interference of complex amplitudes in a high-dimensional vector space to transform the input states with superposition into a solution.

Because of a large number of already existing gate-based optimization algorithms, in this article, we have focused on the linear systems problem (LSP) as one of the most popular and ubiquitous methodologies associated with optimization problems (Dantzig, 2002). The quantum algorithm for the linear systems of equations (QALS), also called the Harrow–Hassidim–Lloyd (HHL) algorithm, is a fundamental quantum algorithm for computing the expectation values in the solution of linear systems (e.g., linear programming and quantum machine learning). Solving linear systems on their own or as a subroutine in more complex problems is a fundamental building block of many optimization algorithms (Harrow et al., 2009). The classical problem of solving a system of linear equations can be described as follows:

$$A \vec{x} = \vec{b} \qquad (7)$$

where $A \in \mathbb{C}^{N \times N}$ denotes the invertible coefficient matrix, $\vec{x} \in \mathbb{C}^N$ indicates a vector of unknowns, and $\vec{b} \in \mathbb{C}^N$ represents the vector of solutions.

Gaussian elimination solves a linear system in the order of $\mathcal{O}(N^3)$.[4] One of the best classical approximative methods (conjugate gradient) returns $\tilde{x}$ such that $\|\tilde{x} - x^*\| \leq \varepsilon$, where $x^*$ is the solution in time $\mathcal{O}\left(N\kappa \log\left(\frac{1}{\epsilon}\right)\right)$ with $N$ = number of rows or columns of the sparse matrix $A$, $\kappa$ = condition number of $A$ (ratio of the largest eigenvalue $\lambda_{max}$ to the smallest eigenvalue $\lambda_{min}$), and $\epsilon$ = target accuracy (see Table 1). Upon the application of the HHL algorithm, this can be improved to the running time of order $\mathcal{O}\left(\frac{\kappa^2 \log(N)}{\epsilon}\right)$. However, it must be taken into account that the exponential runtime improvement[5] of the HHL algorithm is only possible if the original problem (12) is modified to fulfill certain requirements. Therefore, the classical representation is transformed into a quantum state representation with $A|x\rangle = |b\rangle$, constructing a solution quantum state $|x\rangle = A^{-1}|b\rangle$ of expectation values in the solution of the system. Calculating the actual vector $\vec{x}$ would take time, which is, at least, linear in $N$. The necessary modifications and requirements are that these quantum states can be created efficiently $\mathcal{O}(poly(\log(N))$, where $A$ is Hermitian ($A^\dagger = A$) and sparse (polylog $N$ nonzero entries per row/column and entries computed efficiently) (Harrow et al., 2009). A later modification of the algorithm by Ambainis (2012) showed a better scaling in the condition number and a worse scaling in precision with $\mathcal{O}\left(\frac{\kappa \log(N)}{\epsilon^3}\right)$. Further improvements and a complexity comparison between classical and quantum LSP solution algorithms are depicted in Table 1.

---

[3] These are the controlled-Not, Hadamard, and phase-gate $R_{\Pi/4}$.
[4] This is because the usual measurement of complexity $\mathcal{O}(f(N))$ is based on an asymptotic upper bound of $f(N)$ scaling of runtime by the number of elementary operations used by an algorithm.

[5] To date, no mathematical proof exists of the exponential speedup in optimization problems using quantum algorithms. This can currently only be proven heuristically.

**Table 1.** Complexity comparison between classical and quantum LSP solution algorithms (Bernal Neira, 2021)

| Solution algorithm | Order | Subroutines used |
|---|---|---|
| Gaussian elimination | $\mathcal{O}(N^3)$ | RR |
| Conjugate gradient | $\mathcal{O}\left(N\kappa \log\left(\frac{1}{\epsilon}\right)\right)$ | MVM |
| Harrow, Hassidim, & Lloyd (2009) | $\mathcal{O}\left(\frac{\kappa^2 \log(N)}{\epsilon}\right)$ | HS, AA, PE |
| Ambainis (2012) | $\mathcal{O}\left(\frac{\kappa \log(N)}{\epsilon^3}\right)$ | HS, PE, VA |
| Childs, Kothari, & Somma (2017) | $\mathcal{O}\left(\kappa \log(N)\, poly \log\left(\frac{1}{\epsilon}\right)\right)$ | HS, VA, LU |
| Subasi, Somma, & Orsucci (2019) | $\mathcal{O}\left(\frac{\kappa \log(N)}{\epsilon}\right)$ | HS |
| An & Lin (2022) | $\mathcal{O}(\kappa\, poly(\log(\kappa/\varepsilon)))$ | HS |

RR = row reduction; HS = Hamilton simulation; AA = amplitude amplification; MVM = matrix–vector multiplication; PA = phase estimation; VA = variable–time amplitude amplification; LU = linear combination of unitaries

Operator $A$ functions as a (local) Hamiltonian describing the time evolution with $e^{-iAt}$ acting on a state $|b\rangle$ for the superposition of different time $t$. Matrix $A$ can be developed into a basis with eigenstates $|u_j\rangle \in \mathbb{C}^N$ and the respective eigenvalues $\lambda_i \in \mathbb{C}$ as follows:[6]

$$A = \sum_{j=1}^{N} \lambda_i |u_j\rangle\langle u_j|, \text{ and } A^{-1} = \sum_{j=1}^{N} \lambda_i^{-1}|u_j\rangle \qquad (8)$$

The quantum state $|b\rangle$ can also be expanded into the same eigenstate basis with the following:

$$|b\rangle = \sum_{j=1}^{N} \beta_j |u_j\rangle, \quad \beta_j \in \mathbb{C} \qquad (9)$$

We can decompose $|b\rangle$ in the eigenstate basis of $A$ to estimate the corresponding eigenvalues $\lambda_i$. With the application of a quantum phase estimation (QPE), the eigenvalues of $A$ can be estimated using a quantum Fourier transform (QFT) (for more detailed descriptions, see Nielsen & Chuang, 2010). QFT generates a basis change from the computational basis $\{|0\rangle, |1\rangle\}$ to the Fourier basis $\{(|0\rangle + |1\rangle)/\sqrt{2}, (|0\rangle - |1\rangle)/\sqrt{2}\}$. The solution quantum state can be now calculated as follows:

$$|x\rangle = A^{-1}|b\rangle = \sum_{j=1}^{N} \frac{\beta_j}{\lambda_j} |u_j\rangle \qquad (10)$$

---

[6] We are not using $|e_i\rangle$ for the eigenstates and $E_i$ for the respective eigenvalues as done in (8) to account for the fact that the considered state

The implementation of the HHL algorithm can be split into three subroutines (see Fig. 5).

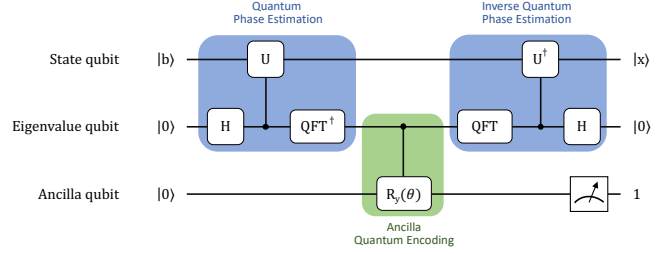

**Fig. 5.** HHL Algorithm

## 2.3 QUANTUM ANNEALING VERSUS GATE-BASED OPTIMIZATION ALGORITHMS

In general, quantum annealing can solve a large class of combinatorial optimization problems with a huge variety of important real-world applications (Symons et al., 2023). However, it is only able to directly solve Boolean optimization problems (like QUBO) that does not represent universal quantum computation (Parekh et al., 2016). The generic algorithm can be applied to any constraint satisfaction problem (CSP) and processed by a quantum computer when it is mappable in its topology. It utilizes the natural tendency of quantum systems to relax into the state of least energy, which requires less control over qubits. QA is based on the adiabatic theorem that requires no heat transfer between a thermodynamic system and its environment (see Section 2.1). Real technical implementations always ultimately lead to some, albeit very small, energy exchange. Even for quantum computers operating in an ultracold temperature range ($T <$ 15mK), finite heat exchange only ever leads to a near-optimal solution. In general, the finite annealing temperatures must decrease with increasing problem size (Albash et al., 2017). The smaller the spectral minimum gap $\Delta E_{min}$ is, the greater is the probability that the best solution will not be achieved as the lowest-energy state. This determines the minimal run time for an adiabatic process with $\mathcal{O}(\Delta E_{min}^{-2})$ (see Section 2.1). The goal to find the ground state with high accuracy as quickly as possible, requires a sufficiently slow change of the energy states so that the evolution time might by exponential (Farhi et al., 2008). Therefore, a current performance evaluation with numerical experiments is only possible for a smaller problem size, while larger practical problems are at present no longer feasible. There is a delicate balance between the annealing time and the expected solution quality, so the optimal annealing time for a problem is not fixed a priori (Parekh et al., 2016). Users should be able to comprehensively evaluate this quality vs time trade-off to assess the total cost of QA optimization in a causal manner (Lubinski et al., 2023). This is the reason why unlike other

---

space does not consider energies here but solutions for the linear equation system in (12).

quantum algorithms, QA lacks general, rigorous worst-case upper bounds on its runtime (Montanaro, 2016).

Compared with those of quantum circuits, the advantages of quantum annealing are that the technical realization of the relevant analog computer hardware is simpler and the number of qubits can be scaled up more easily. QA is more tolerant against environmental noise and therefore less prone to errors than gate-based computers. Additionally, QA can solve problems considerably more efficiently than a gate-based algorithm. However, this efficiency is bought about by the restriction in the breadth of the application, which is limited to optimization problems (mainly combinatorial optimization) and sampling[7] only (e.g., the Shor algorithm cannot be implemented with QA). Future applications of QA depend on the scalability of noisy qubits to solve larger problems.

In contrast to QA, where optimization processes are automatic, in the gate-based approach, quantum state evolution can be controlled and manipulated. The advantages of gate-based quantum computing are a broader application area than QA and a variety of technical implementation options (see Fig. 1 – right column). Already developed powerful algorithms, (e.g., Shor and Grover) allow one to solve a bigger class of problems. As optimizations can be understood as discredited adiabatic computation, QA tasks can also be executed. Keever & Lubasch (2023) have presented classical algorithms for the optimization of gate-based quantum circuits for adiabatic quantum computations, which can significantly outperform the Trotter product formulas.

Gate-based quantum computing requires a higher effort in incorporating algorithms and hardware design, as logical variables are mapped to multi-state qubits. Theoretical quantum gate models assume error-free qubits, error-free quantum gates and arbitrary connectivity. Today's quantum computers are determined by the loss of quantum coherence due to environmental interaction and quantum noise, which tends to destroy the fragile quantum state (Schaller et al., 2006). As the problem size increases, the gate accuracy decreases and the overhead cost of performing quantum error correction increases significantly (Preskill, 2018). Consequently, reliability and coherence time requirements are a critical issue. The noise in quantum gates that restricts the size of quantum circuits will limit the size of quantum circuits that can be executed reliably (see Section 4).

**2.4 APPLICATION AREAS OF QOAS IN OR**

Optimization algorithms based on quantum computing have wide applications in OR. The recent extreme increase in the number of publications in the field of QOA shows its growing importance. The following is a brief list of six main OR application areas that are highly likely to be affected by the use of quantum optimization algorithms in the future. With some current application examples, a short overview shall be given without claiming to be a complete and well-founded literature review:

**2.4.1 SEQUENCING**

Quantum algorithms have been used in product sequencing to optimize picking processes in warehouses and distribution centers (Atchade-Adelomou et al., 2021). A QUBO algorithm (see Section 2.1) was used to control a picking robot to minimize the total transportation distance. The goal was to efficiently manage picking orders while minimizing the picking robots used.

Yarkoni et al. (2021) presented an automotive industry application of a quantum annealing approach to solve a binary (two-color) paint shop problem. The application showed that the used Ising model (affine problem to QUBO) is well-suited for smaller problems of the NP-hard problem of minimizing the number of color switches between bodies in a paint shop queue during manufacturing. For large problem sizes, the performance of the quantum algorithm quickly approaches that of a simple greedy algorithm. Huang (2022) solved the same combinatorial problem by using a quantum algorithm for the approximate optimization for multi-color changes in a paint line of an automobile factory. Three vehicles were used to demonstrate how sequencing problems can be formulated and solved with quantum algorithms. Future extensions of the model will require an increase in the number of vehicles with multiple paint lines. This can be achieved by using noise reduction techniques as well as further adaptations of quantum algorithms (Streif et al., 2021).

**2.4.2 ROUTING**

Typical applications of quantum algorithms in routing are to optimize vehicle transport routes, fleet size, and transportation costs combined with lower emissions. Traffic management also represents an important area of application. QOA application supports smart cities with efficient public transport by optimizing fleet deployment and routing, while minimizing waiting times for passengers and reducing air pollutions and accidents.

A bi-objective inventory routing model with the objectives to reduce emissions and to minimize the fleet size and inventory costs was introduced by Alinaghian and Zamani (2019). Augerat's test problem set has been used to generate small and large problem instances (Augerat, 1995). A meta-heuristic genetic quantum algorithm (GQA) with constant and random selection strategies has been applied to this NP-hard problem. A comparison of the results with a conventional non-dominated sorting genetic algorithm (NSGA) shows the better performance of the proposed quantum algorithm in terms of the defined Pareto criteria.

Ma et al. (2021) solved an NP-hard split demand vehicle routing problem (SDVRP) with minimal vehicles and avoiding task splits with a GQA. A transportation simplex method is incorporated with GQA to transform the observed solutions into real loading schemes. The quantum algorithm uses a simple nearest-neighborhood-based heuristic to generate vehicle routes and adopts a local search method to improve solution quality. The results show that the combined quantum algorithm splits few tasks and can generate many solutions better than the capacitated vehicle routing problem (CVRP) best known in the TSPLIB 95 sample instances.

---
[7] Sampling problems are related but slightly different to optimization problems where not one but a number of low energy states are sampled in order to evaluate an energy landscape.

**Table 2.** Application areas of QOAs in OR

| Application area | Decision problem | Sector | References |
|---|---|---|---|
| Sequencing | • Picking order sequencing<br>• Multi-car paint shop sequencing | Retail<br>Automotive | Atchade-Adelomou et al. (2021)<br>Yarkoni et al. (2021); Huang et al. (2022); Streif et al. (2021) |
| Routing | • Inventory routing problem (IRP)<br>• Split demand vehicle routing problem (SDVRP)<br>• Real-time bus fleet navigation<br>• AGV route optimization<br>• Capacitated vehicle routing problem (CVRP)<br>• Traffic flow forecasting | Green distribution<br>Transport<br>Bus transport<br>Manufacturing<br>Transport<br>Smart cities | Alinaghian & Zamani (2019)<br>Ma et al. (2021)<br>Yarkoni et al. (2020)<br>Li et al. (2020)<br>Feld et al. (2019)<br>Zhang et al. (2021) |
| Scheduling | • Parallel flexible job shop scheduling (PFJSP)<br>• Permutation flow shop problem (PFSP)<br>• Runway scheduling<br>• Scheduling of port resources<br>• No-wait flow shop scheduling (NWFSP)<br>• Flight-gate assignment problem<br>• Load time interval scheduling | Manufacturing<br>Manufacturing<br>Airport operations<br>Port operations<br>Manufacturing<br>Aviation<br>Charging of EVs | Denkena et al. (2021)<br>Chen et al. (2020)<br>Tran et al. (2016)<br>Su et al. (2020)<br>Zhu et al. (2019)<br>Chai et al. (2023), Stollenwerk et al. (2019)<br>Dalyac et al. (2021) |
| Batching and Loading | • Lot sizing<br>• Aircraft cargo loading | Manufacturing<br>Aviation | Zhang & Wang (2018)<br>Nayak & Sahu (2022); Pilon et al. (2021) |
| Inventory Management | • Inventory control problem<br>• Newsvendor problem | Warehousing<br>Operations management | Jiang et al. (2022)<br>Gacon et al. (2020) |
| Finance | • Portfolio optimization<br>• Risk/crash prediction<br>• Pricing | Investment management<br>Risk management | Kerenidis et al. (2019); Slate et al. (2021)<br>Orus et al. (2019); Woerner & Egger (2019); Egger et al. (2020); Miyamoto (2022)<br>Stamatopoulos (2022); Miyamoto & Kubo (2021) |

For the 2019 WebSummit Conference in Lisbon, Volkswagen partnered with the city of Lisbon for a pilot project to apply quantum algorithms for traffic control (Yarkoni et al., 2020). In the first phase, the movement of people from previous conferences was analyzed to build bus routes throughout the city. In the second phase, a navigation app installed on nine buses on three routes was used in combination with a quantum annealer from D-Wave, connected to live traffic data, to optimize bus routes in near real-time (updating every 2 min). The exemplary results showed that critical traffic situations could be bypassed, thus reducing the impact of the conference on city traffic caused by this.

Other interesting routing application areas for QOA are automated guided vehicle (AGV) path planning based on quantum ant colony optimization (Li et al., 2020), a capacitated vehicle routing problem (Feld et al., 2019), and traffic flow forecasting, based on quantum particle swarm optimization (Zhang et al., 2021). Osaba et al. (2022) identified 53 different papers in the intersections of quantum computing and routing problems.

### 2.4.3 Scheduling

Denkena et al. (2021) presented a quantum annealing-based optimization for process-parallel flexible job shop scheduling. A realistic use case demonstrated the good performance and practicability. Although the results were similar to those of classical heuristics, the new approach delivered reliably low annealing times for large problem instances and was robust against stochastic influences.

A quantum-inspired ant colony algorithm (QIACO) was used by Chen et al. (2020) to optimize a two-stage permutation flow shop with batch processing machines (BPMs). With the consideration of different order sizes and arbitrary arrival times, the makespan was minimized. A benchmark with the classical hybrid discrete differential evolution (HDDE) and batch-based hybrid ant colony optimization (BHACO) algorithms shows the advantages of the QIACO in terms of both solution quality and running time.

Tran et al. (2016) presented a hybrid quantum–classical approach to solving airport runway scheduling. A quantum annealer that samples from the configurations space is combined with a classical processor that maintains a global search tree and enforces constraints on the relaxed components of the problem. Empirical results showed that QA search can be used for more effective pruning of search nodes and better heuristics for node selection than a standard classical approach.

Other interesting scheduling application areas for QOA are the prediction of vessel traffic volume to the rational scheduling of port resources (Su et al., 2020), a quantum-inspired cuckoo co-search algorithm to minimize the makespan in a no-wait flow shop scheduling problem (Zhu et al., 2019), flight gate assignment problems (Stollenwerk et al., 2019; Chai et al., 2023), and the optimal scheduling of the load time intervals of electric vehicles (Dalyac et al., 2021).

### 2.4.4 Batching and loading

Zhang and Wang (2018) applied a quantum evolutionary algorithm to the lot-sizing planning problem. A quantum spin angle was used to control the gene mutation rate and maintain the genetic information that the optimal individuals involved. According to a case study, this hybrid approach combining a quantum algorithm with a classical evolutionary genetic algorithm showed higher resolution accuracy and convergence rate than the corresponding classical counterparts, to effectively solve complex multi-constrained batching problems.

Kayak and Sahu (2022) solved an aircraft loading optimization problem that simultaneously considered maximizing the aircraft payload capacity and minimizing the loading and unloading operations during multiple stopovers to reduce operating costs. Both annealing-based and gate-model quantum computing have been used to solve the problem. Annealing-based computing has been found to be more promising as it provides a sufficient number of qubits to address a realistic problem size. Similar results were described earlier in the Airbus Quantum Computing Challenge. Pilon et al. (2021) used a QUBO algorithm for benchmarking different classical solvers. A QBsolv solver for QUBO functions was used on the basis of the heuristic Tabu search. It has been shown that the quantum-inspired solver is efficient to validate the model and agrees with the results of the exact solver for small problem sizes. However, for medium and large problem sizes, it is difficult to estimate the problem solution quality.

### 2.4.5 Inventory management

Jiang et al. (2022) developed a quantized policy iteration algorithm for a Markov decision process to solve an inventory control problem. With the application of an HHL algorithm (see Section 2.2), the practicality of quantum algorithms to solve small inventory control problems was shown. This algorithm has been predicted to be practical in the long term with the availability of fault-tolerant quantum computing technologies (see Section 3). However, in the near term, variational analogs of HHL with similar capabilities are the focus of application.

A newsvendor problem was solved by Gacon et al. (2020) by applying a quantum-enhanced algorithm for simulation-based optimization (QSBO). QSBO uses quantum amplitude estimation (QAE) to offer a quadratic speedup for the evaluation of the objective function as compared to the classical Monte Carlo simulation for both continuous and discrete variables. It has been shown that QSBO is particularly well suited for objective functions defined as the expectation value, variance, or cumulative distribution function.

### 2.4.6 FINANCE

The financial sector has historically been an early adopter of quantum computing by investing in research and development efforts in the area of quantum finance (Dalzell et al., 2023). One major application in finance is the portfolio optimization problem to find the optimal asset investment allocation to maximize expected returns with bearable market risks. Kerenidis et al. (2019) developed the first quantum algorithm for the constrained portfolio optimization problem. With a moderately accurate solution, the applied quantum interior-point method (IPM) can achieve a polynomial speedup over the best classical algorithm. A further highly efficient quantum walk optimization algorithm (QWOA) for portfolio optimization was proposed by Slate et al. (2021). A discrete mean-variance Markowitz model was used to provide numerical evidence for the efficiency of QWOA in NP-hard portfolio optimization with discrete asset constraints. The use of QWOA has been shown to reduce the search space by a significant factor, resulting in consistently improved performance in obtaining a high-quality portfolio configuration with fewer iterations and with significantly lower standard deviations in numerical simulations.

A further important application of QOA is risk and crash prediction on financial markets. By applying a QUBO algorithm, Orus et al. (2019) mapped the equilibrium condition of a toy-model financial network to the ground-state problem of a quantum annealer (see Fig. 3). The results indicated a potentially more efficient way to assess financial equilibrium and predict financial crashes. Woerner and Egger (2019) presented a QOA that analyzes financial market risk more efficiently than traditional Monte Carlo simulations. Risk measures such as value at risk and conditional value at risk have been used on a gate-based quantum computer. For slowly increasing the circuit depths. the algorithm provides a near quadratic speedup with an improved convergence rate as compared to Monte Carlo methods. Egger et al. (2020) and Miyamoto (2022) applied quantum algorithms for calculating risk contributions in a credit portfolio.

Stamatopoulos et al. (2022) introduced a quantum gradient estimation algorithm to compute and assess derivatives. Through QAE, the computation of derivative pricing and Greeks, as associated market sensitivities, have been quadratically accelerated in the target error. The applied quantum algorithm suggests that an additional quantum advantage is possible in risk analysis, on top of the quadratic speedup of derivative pricing. Miyamoto and Kubo (2021) proposed a quantum algorithm for the finite-difference method (FDM) for solving the Black–Scholes partial differential equation.

### 2.5 COMPARATIVE EVALUATION OF THE SOLUTION QUALITY OF QOAS IN OR

Despite the limited number of use cases described in Section 2.4, certain insights can be derived from the comparison of the solution results. It can be seen that the problems described are either based on standardized test problem sets (e.g. Augerat 1995) or were applied to very limited exemplary problems (Yarkoni et al., 2020). The application areas described all have small problem sizes that range between toy models, e.g. of a financial network (Orts et al., 2019) and smaller application models, such as inventory control problems (Jiang et al. 2022) or sequencing picking processes (Atchade-Adelomou et al., 2021). It has been shown in many cases that for large problem sizes, the performance of the QOA quickly approaches that of classical algorithms e.g. greedy algorithms (Yarkoni et al., 2021).

A summarized comparison of the surveyed works in Section 2.4 shows that the QOAs used only offer modest speed ups over their classical counterparts. All of the examples described show a maximum quadratic speedup - or polynomial speedup in general - which, however, with a future increase in the problem size can be significant (see Section 3). To date, no QOA with known exponential speedup exists, so that the focus in the development of new QOA is on a faster-than-quadratic speedup.

Despite the lack of exponential runtime improvements, there are currently other performance benefits from applying QOA in OR. In the examples described above, these range from an improved Pareto distribution criteria (Alinaghian and Zamani, 2019), robustness against stochastic influences (Denkena et al., 2021), to an effective pruning of search nodes and better heuristics for node selection (Tran et al., 2016) to a higher resolution accuracy and convergence rate than the corresponding classical approaches (Zhang and Wang, 2018).

In general, it can also be seen that the selection of the classic algorithms used in Section 2.4 inevitably influences the benchmarking results. Powerful simple randomized algorithms, that employ a degree of randomness as part of its logic or procedure, were not used in the performance comparison (Motwani and Raghavan, 1995). It has been shown that randomized classical algorithms for many generic computationally intensive problems without physical structure can work just as efficiently as corresponding quantum routines (Tang, 2019). This is particularly true for applications in the field of finance and machine learning.

### 3. IMPLICATIONS OF APPLYING QOAS IN OR

Even though operations research is just at an early stage for concrete applications of QOAs, initial results and trends are crystallizing for future deployments. The resulting main implications for applying QOA algorithms in OR have been summarized in the form of five implications.

***Implication 1.*** *With the use of quantum computers and the solving of decision problems using QOA, the set of computable decision problems does not change.*

Turing machines, which are a fundamental model of computability and theoretical computer science, can be used to compute an enormous variety of functions. The Church–Turing thesis states that the class of functions that can be computed by a Turing machine corresponds exactly to the class of functions that can be computed by an algorithm (Church, 1937). It turns out that the Church–Turing thesis remains valid even when quantum computers are used. Yao (1993) confirmed that any function that can be computed by a quantum Turing machine in polynomial time can also be computed by a quantum circuit of polynomial size. Quantum computers obey the Church–Turing thesis, so the quantum circuit model (see Section 2.2) can efficiently compute any algorithmic process and any realistic model of classical computation (Mosca, 2008). This quantum extended Church–Turing thesis states that the class of problems that can be computed on a quantum computer is no different from the class of problems that can be computed on a classical computer. This means that quantum computers do not solve decision problems that are intractable for classical computers. The difference between classical and quantum computers lies in the efficiency with which the computation of the decision problem can be performed. This is based on the superposition state of quantum bits, which can be simultaneously in any one of a potentially infinite number of states. Therefore, a linear number of algorithmic operations can produce an exponentially large superposition of states, and an exponentially large number of operations can be processed in parallel in a single step. Hence, some practically important problems can be solved considerably more efficiently on quantum computers than with any known classical algorithm executed on the best classical computer.

***Implication 2.*** *QOAs have the potential to provide faster and more efficient solutions for optimization problems with reduced memory requirements.*

Exponential quantum acceleration is generally more difficult to demonstrate in quantum optimization than in the other two major application areas of quantum simulation and quantum cryptography. In order to perform a better evaluation and comparison of performances, it is necessary to define various quantum speedup categories. These range from provable quantum speedup (e.g., Grover and Shor), strong quantum speedup (compared with the best classical algorithm), and potential quantum speedup (compared with a set of algorithms) to the limited quantum speedup over classical counterparts (Rønnow et al., 2014).

The use and application of quantum algorithms, as compared to that of classical (non-quantum) algorithms, is considerably linked to the computer platform used in each case. Both are mutually dependent and thus closely intertwined. In order to comprehensively analyze the advantages and disadvantages of different QOAs, the existing algorithms are described in the context of the quantum computer architecture used.

Moreover, solving decision problems in the field of operations research requires considering not only the performance side but also the costs incurred by the use of quantum algorithms. All process-relevant cost components must be taken into account: the tracking of classical pre- and post-processing, the explicit instantiation of quantum oracles and data access structures, and ideally the computing of the constant factors of all quantum subroutines (Dalzell et al., 2023). With the assumption of the usual quadratic speedups (e.g., Grover, 1996) by the quantum algorithm of *n* classic bits, a comparison of the quantum processing cost $c_{qbit} \sqrt{n}$ with the classical processing cost $c_{cbit} n$ for the instance size can be expressed as follows:

$$n > \left(\frac{c_{qbit}}{c_{cbit}}\right)^2$$

In accordance with the enormous economies of scale that classical computing has been able to realize over the last decades with the help of Moore's law, this results in a very high quantum–classical cost ratio. The conservative assumption of $10^8$ for this coefficient results in a value of $10^{16}$ qubits for the break-even point of the problem size.

The different uses of various quantum technologies lead to a very heterogeneous cost structure. While photonic systems, for example, still have a laboratory character, the first approaches to industrial production are already emerging in the field of superconducting circuits (see Fig. 1). A comprehensive performance comparison must take into account not only the processing costs but also the number of operations required and the corresponding operation time. Quantum computers need fewer operations, but each operation is slower. The question of quantum advantage can therefore only be answered in the complex interplay of many performance and cost criteria. Bova et al. (2023) showed that quantum computers can still generate economic value via lower cost and lower power consumption even when they do not provide a quantum advantage over classical computers.

***Implication 3.*** *There is good evidence (it is strongly believed) that even quantum algorithms cannot solve NP-complete problems such as the traveling salesman problem (TSP) efficiently in polynomial time.*

Complexity theory research enables a better understanding of which quantum advantages are theoretically possible. To evaluate the performance of quantum algorithms, it is common to use a collection of computational problems with similar difficulties and characteristics, which are grouped into classes. According to these computational complexity classes, some optimization problems are easy to solve in polynomial time (P) as the size of the problem increases, as compared to NP problems where an algorithm can provide a short answer for very hard problems that can be verified in polynomial time (see Fig. 6). Decision problems that can be solved using a polynomial amount of space, belong to the class PSPACE. The subclass NP-complete problems are of special interest, as an algorithm to solve a specific NP-complete problem can be adapted to solve any other problem in NP, with a small overhead (Nielsen & Chuang, 2010). The main question is whether all of the problems have efficient algorithms, where every problem whose solution can be quickly verified can also be solved quickly, better known as the still unproven P = NP problem. The complexity class BQP (which stands for bounded-error quantum polynomial time) is a collection of decision problems that can be efficiently solved in polynomial time on a quantum computer with a bounded error (at most 1/3). With a classical computer, this class of problems can only be solved either deterministically or probabilistically in exponential time. Despite recent results demonstrating quantum supremacy for certain very specific problems,[8] NP-complete problems[9], such as the TSP, are likely to remain quantum hard, and quantum computers are not expected to be able to solve them within practical time constraints.

In general, it is true that $P \subseteq BQP$, and thus, any problem solvable on a classical computer is tractable on a quantum computer. It is widely believed that $NP \nsubseteq BQP$, whereby the proof is extremely challenging (Childs & van Dam, 2010).

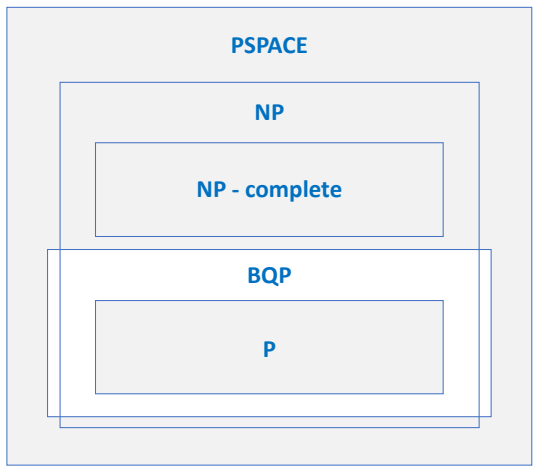

**Fig. 6.** Euler diagram complexity classes

***Implication 4.*** *The creation of QOA is highly non-trivial and most likely to apply to certain problems only.*

Although the application of classical algorithms to quantum computers can mean an acceleration of the solution process, the main focus is on the development of new quantum algorithms. The potential of powerful quantum parallelism for solving optimization problems can only be realized in conjunction with sophisticated QOAs. Intrinsic quantum parallelism is considered to be a necessary condition. As a sufficient condition, interference effects are necessary for an intelligent algorithm design. The basic idea of any quantum algorithm is to strengthen the correct problem solution by the constructive interference and to weaken the wrong solutions by destructive interference. Currently, only a very limited number of QOAs exist, mostly based on simplified, specific problems. Since Shor's algorithm only a handful of fundamental quantum algorithmic kernels, generally providing modest polynomial speedups over classical algorithms, have been invented (Parekh et al., 2019). Solving real-world OR problems requires a generalized QOA that will work for arbitrary, complex functions. Developing a new QOA is very challenging where the algorithm is expected to achieve exponential speedup. In general, quadratic quantum advantages seem generically possible, while larger improvements in reducing the number of elementary arithmetic operations require special structures in the underlying problem (Martyn et al., 2021). Machine learning and artificial intelligence can support this future task (Dunjko & Briegel, 2018). The application of algorithms is strongly linked to the used quantum computer. It is therefore important to distinguish between a quantum algorithm for ideal abstract, physically-inspired abstract and physical quantum computers (Parekh, 2020). Quantum computer hardware constraints present significant challenges for QOA (Coffrin, 2019). Optimization algorithms often demand operations between all pairs of qubits, and thus, the connectivity of the computer hardware will not necessarily match the connectivity of the optimization problem. The potential of quantum computers is restricted by extremely vulnerable quantum hardware, where tiny disturbances can destroy the decoherence needed (noise).

From the point of view of practical OR applications, it is quite often more important to find a good instead of a perfect solution. Even minuscule improvements in optimization can translate to significant savings in resources and drive a tremendous amount of real-world value. Quantum approximate optimization algorithms (QAOAs) tailored to run gate-based devices are of particular interest (Farhi et al., 2014). QAOA aims to efficiently produce a near-optimal solution with a mathematically provable bound on quality. Although currently only some theoretical limitations of the performance of QAOA are known, QAOA might be able to achieve better approximations than possible classically for quantum optimization problems, such as quantum max-cut (Parekh, 2023). Further improvements can be achieved by introducing Variational Quantum Eigensolver (VQE). Amaro et al. (2022) showed that max-cut problems can be solved faster and more reliably than standard QAOA using a combination of VQE and filter operators.

---

[8] The Shor algorithm has shown that factoring and the extraction of a discrete logarithm can be efficiently solved in quantum polynomial time.

[9] This refers to the set of problems that are traditionally conjectured to produce instances too hard for classical computers to solve exactly and deterministically within polynomial time constraints.

*Implication 5. QOA often unfolds its power in mutual interaction with classical computing.*

By combining the conceptional and physical advantages of classical and quantum computing, we can realize future advantages in decision problem solving. The study of the workings and mechanisms of quantum algorithms has led to the development of quantum-inspired algorithms for classical computers. Quantum-enhanced algorithms can simulate quantum systems on classical computers. On the one hand, it may be a classical algorithm emulating a quantum phenomenon to achieve faster solutions. This has led to new research directions such as quantum machine learning. On the other hand, it may be the use of existing classical algorithms to improve quantum optimization. Wang (2022) showed that a classically boosted quantum optimization algorithm (CBQOA) can solve a wide range of combinatorial optimization problems, including all unconstrained problems and many important constrained problems. An important advantage of CBQOA is an efficiently implemented continuous-time quantum walk on an appropriately constructed graph that connects the feasible solutions.

Further advantages result from the hybrid interaction between classical and quantum computing. Before fully fault-tolerant quantum computers can be realized in the future (see Section 2.3), noisy intermediate-scale quantum devices (NISQs) offer a current deployment alternative (Preskill, 2018). NISQ devices are near-term quantum computers, that are not yet sufficiently advanced for significant error correction, longer coherence times, and improved noise control. NISQ uses a hybrid concept with only a limited number of qubits, that combines quantum devices with classical computers. The standard quantum algorithms such as Shor and Grover are not suitable for the NISQ era, because they need many qubits with a very high degree of accuracy. A positive effect of NISQ hardware is that it provides a unique opportunity to develop and numerically evaluate novel quantum heuristic methods, like the variational quantum algorithms (VQAs), which use a classical optimizer to train a parameterized quantum circuit (Cerezo et al., 2021).

In general, quantum computers cannot work efficiently without the assistance of classical computers. The aim is not for quantum computers to completely replace conventional computers. If possible, modern classical algorithms should be used and only highly specialized subroutines with a quantum advantage should be executed as quantum algorithms. In addition, classic computers are used for the pre- and post-processing of data. The aim is therefore to solve OR optimization problems in the future with the smallest possible quantum footprint.

**4. CONCLUSION**

Quantum computing opens up new interesting possibilities for the development and use of optimization algorithms in operations research. QOAs promise to solve decision problems with larger problem spaces and provide better quality solutions and faster solution times. The superposition of qubits allows to maintain and therefore process multiple possible states simultaneously to enable an intrinsic form of parallel algorithms. Solving hard optimization problems in many OR fields with parallel algorithms is crucial to make at least some problem instances tractable in practice (Schryen, 2020).

Although initial results in the application of QOA are promising, many challenges remain to be overcome for future widespread implementation. The demonstration of bounded quantum accelerations for specific tailored problems do not suggest a general quantum advantage for real decision problems in OR. For this, a broader and deeper performance comparison between the classical and the quantum-based optimization algorithms is required. User-oriented benchmark problems are needed under different framework conditions and problem sizes. More OR-related near-term quantum-inspired optimization problems are necessary to complement, validate, and leverage the existing testbed efforts. The problem formulation and configuration of classical algorithms play a crucial role for the benchmarking results. Therefore, the proper selection and configuration of classical algorithms used in quantum benchmark studies is a critical issue (Parekh et al., 2016). Due to the limited computing capacity of quantum computers, optimization problems with large quantum advantages should be selected, which are already evident in small, implementable instances. At the same time, benchmarks should be accessible to users outside of academia and quantum computing developers, they must be integrated with emerging quantum computing benchmark suites and results presented in a manner that is meaningful to experts within operations research (Lubinski et al., 2023).

In addition to a more in-depth empirical confirmation of QOA performances, there are hardware-related challenges in its large-scale implementation. Qubits are unreliable and noisy, and thus, algorithmic quantum operations cannot be performed without significant errors. Therefore, logical qubits need additional qubits for error correction, which lowers the maximum performance of the quantum processing unit (QPU). With the future increase in used qubits, the uncontrolled interactions are increasing. A crucial future issue is the question of how noise can be overcome in the absence of error correction and therefore enable any potential quantum advantage (Parekh, 2023). It has been shown that unless the current noise rates are drastically reduced or the computer topology is adapted to the architecture of the problem, significant speedups for QOA will become impossible (Stilck Franca & Garcia-Patron, 2021). Even noise mitigation techniques based on the post-processing of the quantum computing measurement outcomes would not change the actual result quality. Currently, the use of NISQ devices (see Section 3) severely limits future applications of QOA (see implication 4). Considerable research work and technical improvements to existing quantum computer technology (e.g. noise reduction, error correction) are still required for a future broad application of QOAs in operations research.

Stollenwerk, T., Lobe, E., & Jung, M. (2019, February). Flight gate assignment with a quantum annealer. In *International Workshop on Quantum Technology and Optimization Problems*(pp. 99-110). Cham: Springer International Publishing.

Streif, M., Yarkoni, S., Skolik, A., Neukart, F., & Leib, M. (2021). Beating classical heuristics for the binary paint shop problem with the quantum approximate optimization algorithm. *Physical Review A*, *104*(1), 012403.

Su, G., Liang, T., & Wang, M. (2020). Prediction of vessel traffic volume in ports based on improved fuzzy neural network. *IEEE Access*, *8*, 71199-71205.

Subasi, Y., Somma, R. D., & Orsucci, D. (2019). Quantum algorithms for systems of linear equations inspired by adiabatic quantum computing. *Physical review letters*, *122*(6), 060504.

Symons, B. C., Galvin, D., Sahin, E., Alexandrov, V., & Mensa, S. (2023). A Practitioner's Guide to Quantum Algorithms for Optimisation Problems. *arXiv preprint arXiv:2305.07323*.

Tang, E. (2019, June). A quantum-inspired classical algorithm for recommendation systems. In *Proceedings of the 51st annual ACM SIGACT symposium on theory of computing* (pp. 217-228).

Tran, T., Do, M., Rieffel, E., Frank, J., Wang, Z., O'Gorman, B., ... & Beck, J. (2016). A hybrid quantum-classical approach to solving scheduling problems. In *Proceedings of the International Symposium on Combinatorial Search*, *7*(1), 98-106.

Wang, G. (2022). Classically-boosted quantum optimization algorithm. *arXiv preprint arXiv:2203.13936*.

Watrous, J. (2008). Quantum computational complexity. *arXiv preprint arXiv:0804.3401*.

Woerner, S., & Egger, D. J. (2019). Quantum risk analysis. *npj Quantum Information*, *5*(1), 15.

Yao, A. C. C. (1993). Quantum circuit complexity. In *Proceedings of 1993 IEEE 34th Annual Foundations of Computer Science* (pp. 352-361). IEEE.

Yarkoni, S., Alekseyenko, A., Streif, M., Von Dollen, D., Neukart, F., & Bäck, T. (2021). Multi-car paint shop optimization with quantum annealing. In *2021 IEEE International Conference on Quantum Computing and Engineering (QCE)* (pp. 35-41). IEEE.

Yarkoni, S., Neukart, F., Tagle, E. M. G., Magiera, N., Mehta, B., Hire, K., ... & Hofmann, M. (2020). Quantum shuttle: traffic navigation with quantum computing. In *Proceedings of the 1st ACM SIGSOFT International Workshop on Architectures and Paradigms for Engineering Quantum Software* (pp. 22-30).

Zhang, H. and Wang, Ch. (2018). Lot-sizing based on quantum evolutionary algorithm, *Academic Journal of Manufacturing Engineering*, *16*(4), pp 122-127.

Zhang, D., Wang, J., Fan, H., Zhang, T., Gao, J., & Yang, P. (2021). New method of traffic flow forecasting based on quantum particle swarm optimization strategy for intelligent transportation system. *International Journal of Communication Systems*, *34*(1), e4647.

Zhu, H., Qi, X., Chen, F., He, X., Chen, L., & Zhang, Z. (2019). Quantum-inspired cuckoo co-search algorithm for no-wait flow shop scheduling. *Applied Intelligence*, *49*, 791-803.